\documentclass[12pt,a4paper]{article}

\usepackage[british]{babel}

\usepackage[a4paper,top=2cm,bottom=2cm,left=2.5cm,right=2.5cm,marginparwidth=1.75cm]{geometry}

\usepackage[style=apa, backend=biber]{biblatex} % APA 7th edition style citations using biblatex
\DeclareLanguageMapping{british}{british-apa} % Set language mapping
\DeclareFieldFormat[article]{volume}{\apanum{#1}} % Format volume number

\usepackage{amsmath}
\usepackage{graphicx}
\usepackage[colorlinks=true, allcolors=blue]{hyperref}
\usepackage{hyperref}
\usepackage[title]{appendix}
\usepackage{mathrsfs}
\usepackage{amsfonts}
\usepackage{booktabs} % For \toprule, \midrule, \botrule
\usepackage{caption}  % For \caption
\usepackage{threeparttable} % For table footnotes
\usepackage{algorithm}
\usepackage{algorithmicx}
\usepackage{algpseudocode}
\usepackage{listings}
\usepackage{enumitem}
\usepackage{chngcntr}
\usepackage{booktabs}
\usepackage{lipsum}
\usepackage{subcaption}
\usepackage{authblk}
\usepackage[T1]{fontenc}    % Font encoding
\usepackage{csquotes}       % Include csquotes
\usepackage{diagbox}
\usepackage{float}

\usepackage{setspace}
\usepackage{titlesec}
\titleformat{\section} % Redefine section numbering format
  {\normalfont\Large\bfseries}{\thesection.}{1em}{}

\usepackage{lineno} % Add the lineno package

\rightlinenumbers % Right-align line numbers

\usepackage{float}   % for customizing caption position
\usepackage{caption} % for customizing caption format


\title{Adaptive Money Market Interest Rate Strategy Utilizing Control Theory}

\author[1]{Yuval Boneh}
\affil[1]{\small Byte Masons}
\affil[1]{Yuval Boneh: \texttt{yuvi@bytemasons.com}}

\begin{document}
\maketitle

\begin{abstract}
  Decentralized Finance (DeFi) money markets have seen explosive growth in recent years, with billions of dollars borrowed in various cryptocurrency assets. Key to the safety of money markets is the implementation of interest rates that determine the cost of borrowing, and govern counterparty exposure and return. In traditional markets, interest rates are set by risk managers, portfolio managers, the Federal Reserve, and a myriad of other sources depending on the market function. DeFi enables an algorithmic approach that typically relies on interest rates being directly dependent on market utilization. The benefit of algorithmic interest rate management is the system's continual response to market behaviors in real time, and thus an inherent ability to mitigate risks on behalf of protocols and users. These interest rate strategies target an optimal utilization based on the protocol's risk threshold, but historically lack the ability to compensate for excessive or diminished utilization over time. This research investigates contemporary DeFi interest rate management strategies and their limitations. Furthermore, this paper introduces a time-weighted approach to interest rate management that implements a Proportional-Integral-Derivative (PID) control system to constantly adapt to market utilization patterns, addressing observed limitations.
\end{abstract}

\textbf{Keywords}:
Decentralized Finance, money market, control system, utilization, interest rate.

%-------------------------------------------
% Paper Body
%-------------------------------------------
\section{Introduction}

Decentralized Finance (DeFi) is a cornerstone application of blockchain technology, leveraging smart contracts to enable secure and transparent financial transactions without reliance on central intermediaries. Key DeFi activities include crypto-asset trading via decentralized exchanges (DExes) like Uniswap and Balancer, and crypto-asset borrowing and lending, facilitated by platforms such as Aave, Ajna, and Morpho Blue.

Lending platforms operate by allowing lenders to deposit their crypto-assets into liquidity pools of loanable funds hosted by smart contracts. These pools then provide a source of liquidity for borrowers. In order to ensure the protocol remains solvent, borrowers must provide collateral to secure a loan. Since a borrowers' collateral must be able to recover their debt, loans are generally over-collateralized. Depending on the nature of DEx liquidity available to facilitate liquidation, and the anticipated volatility of the respective lend and borrow assets, the parameters that govern the over-collateralization of a loan are crucial for securing lenders' assets. These parameters include Loan To Value (LTV) ratios (the maximum allowable value of a loan compared to the value of the collateral), Liquidation Thresholds (LTs) (the ratio of borrowed value to collateral value that will trigger liquidation), and interest rates (accumulating debt value), to name a few. In most cases, interest rates are algorithmically determined, considering the total supply of a crypto-asset and the amount of the respective asset borrowed. This is referred to as Utilization and describes the available liquidity for borrowers to borrow, or for lenders to withdraw. Therefore, Utilization is intentionally kept below one to grant lenders flexibility to redeem their assets.

The interest rate model (IRM) plays a critical role in ensuring the safety of the protocol and its users, with several different approaches being utilized by different protocols. This paper presents contemporary interest rate approaches utilizing control theory and investigates their respective benefits and limitations in order to understand enduring risks. Understanding these risks, an alternative interest rate strategy utilizing control theory is proposed and developed for on-chain application.

\section{Representing Interest Rate Strategies utilizing Control Theory}
\begin{figure}[!ht]
  \centering
  \includegraphics[width=0.4\linewidth]{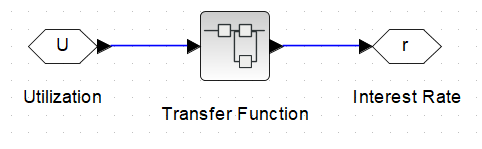}
  \caption{\label{fig:open-loop}Diagramatic representation of a conventional interest rate strategy.}
\end{figure}

Lending market interest rate strategies can be simply represented by an open loop control system, where market utilization provides an input signal, the respective equations perform as a Transfer Function, and the resulting interest rate is the output signal. A simple IRM is presented in Figure \ref{fig:open-loop}.

\begin{figure}[!ht]
  \centering
  \includegraphics[width=0.5\linewidth]{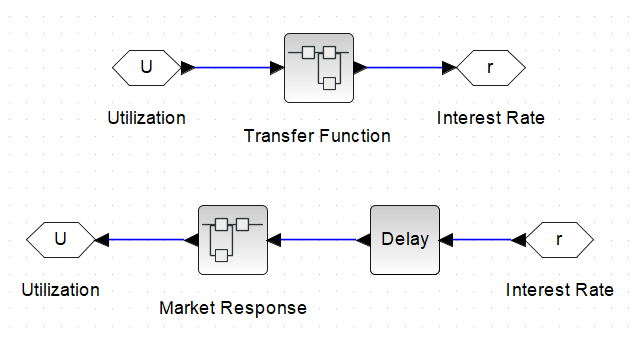}
  \caption{\label{fig:closed-loop}Diagramatic representation of a conventional interest rate strategy with market response.}
\end{figure}

In reality there exists a pseudo feedback loop whereby the market responds to rates and adjusts Utilization. The effective control system is represented in Figure \ref{fig:closed-loop}, however, pursuing accurate modeling of a market response assumes market psychology and is not practical.

The Delay is introduced to avoid a time singularity, as Utilization cannot modify rates at the exact moment that rates lead to a change in Utilization. This approach to modeling identifies the absence of a reference signal, being the optimal utilization rate, and the signal processing associated with more responsive control systems. By comparing a reference signal and processing signal error, a PID controller can be implemented to improve the response of interest rates. Research by Morpho correctly identifies that some derivative of a PID controller is often utilized by decentralized lending protocols, and the justification grounded in reduced computation is sound. The gap between broad implementation and conventional PID control is in correct implentation and signal processing of an error signal.

\subsection{Aave IRM}

The largest DeFi lending market, Aave, employs a piecewise function by setting a base rate and adding to it a linear function of Utilization, that changes gradient beyond some 'Optimal Utilization' \parencite{aaveWhitepaper}. Aave implements transfer functions equal to the piecewise architecutre of the interest rate strategy design, and transfer Function configurations change depending on the market and the risk tolerance, and are described in Aave's documentation \parencite{aaveIrm}. Modeling the Aave Interest Rate Strategy has been previously completed and published \parencite{aaveModel}. The inflection point exists to set a baseline risk-adjusted market rate at a utilization that reflects the volatility of the asset, limiting the protocol's exposure to debt and ensuring that debt can be recovered in the event of liquidations. It works on the notion that if utilization is higher than optimal, interest rates should be high enough to persuade borrowers to repay their loans, lenders to add collateral (majority of the interest paid by borrowers is directed to lenders), or result in liquidation. In PID terms, the Aave system treats zero utilization as the reference and passes the error signal through a unit proportional gain prior to the transfer function.

The main flaw of this approach is that the market's determination of an acceptable interest rate is independent of the strategy's inflection point. This can result in utilization being sustained at dangerously high levels, exposing the protocol to significant levels of debt, and lenders to long-term illiquidity, compensated insufficiently for market conditions. Historically this has been mitigated with human intervention, by modifying the interest rate strategy, issuing incentives, or imposing drastic interest rate measures to force large borrowers to take action. This approach is unsustainable, unpredictable, and may expose the protocol and its users to unnecessary risks.

\subsection{Ajna IRM}

Ajna takes an alternative approach by setting an initial interest rate and increasing or decreasing the rate every 12-hour epoch based on utilization \parencite{ajnaWhitepaper}. If utilization is above the defined 'Target Utilization', then the interest rate is increased by a factor of 1.1. If it is below, then the interest rate is multiplied by a factor of 0.9. Ajna implements a conditional transfer function that switches based on utilization. It concerns itself with whether utilization is equal to or above or below the target, dismissing magnitude, and scaling the interest rate accordingly. Again in PID terms, Ajna treats target utilization as the reference, but uses a normalized proportional gain to create a binary output to modify the interest rate.

This has the benefit of adapting to market conditions by continually increasing or decreasing interest rates based on utilization, however its response time is limited by its design. While this architecture can reasonably mitigate long-tail utilization risks, its response to short term utilization risks is inherently limited by the epoch-bound scaling factor.

\subsection{Morpho IRM}

The AdaptiveCurve interest rate model used by Moprho Blue takes a piecewise linear function similar to Aave's and applies a variable scaling factor equal to the rate of change of the logarithm of the reference interest rate with respect to time (\cite{morphoDocs}). This in turn depends on the error between the actual and desired utilization ratios. \textcite{morphoPaper}  derive the Morpho Blue interest rate function as per Equation \ref{eqMorpho}, which presents a nonlinear PD controller.

\begin{equation}
  \frac{d}{dt}log(r(t)) = k_p * err(u(t)) + \frac{\xi'(u(t))}{\xi(u(t))}u'(t),
  \label{eqMorpho}
\end{equation}

where $k_p$ is a proportional gain factor and is applied to the error signal of the utilization function $u(t)$. $\xi$ is a piecewise function similar to the Aave model.

The PD approach partially mitigates the risks introduced by both previous systems. The system is free to adapt to market conditions, enabling it to actively respond to excessive utilization, and the introduction of the derivative term allows the system to further compensate for the rate of change of utilization. Where the Ajna model requires time to iterate through epochs and scale interest rates, the Morpho model recognizes sharp increases in utilization and compensates the scaling system to account for the change. 

The main inefficiency of the Morpho model is the method's sensitivity to fluctuations around target utilization. Since the system's base interest rate  employs a piecewise function, there is a distinct step in the interest rate gradient at target utilization. Additionally, the speed factor is effectively reset with each interaction, meaning interest rates will aggressively shift from relatively small and decreasing to relatively large and increasing. It is assessed that this system may experience unnecessary interest rate volatility around target utilization, and thus increased settling times.

\section{PID IRM}

\begin{figure}[!ht]
  \centering
  \includegraphics[width=0.8\linewidth]{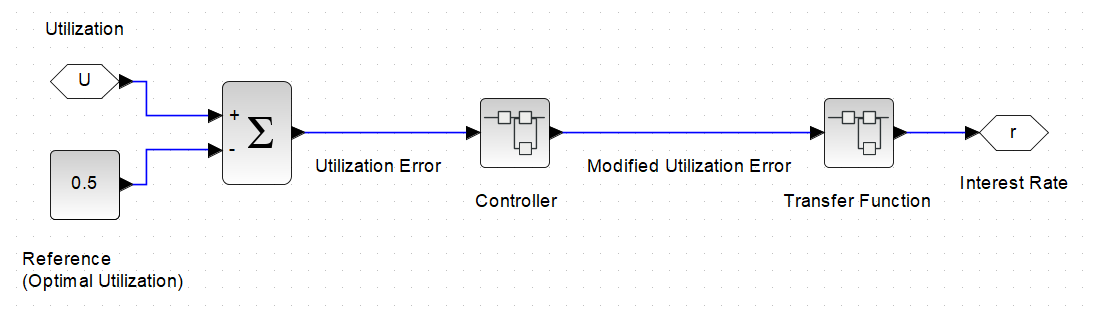}
  \caption{\label{fig:pid-irm}Diagramatic representation of the interest rate strategy model.}
\end{figure}

The strategy explained herein aims to address these flaws by utilizing PID control theory to construct a novel interest rate strategy with a generic interest rate curve that responds to a modified utilization signal, adjusting rates based on the 'Utilization Error'. The premise is that should the market have an appetite to accept higher interest rates, utilization will increase beyond optimal, and the controller will accumulate error, which will grow interest rates further, until the market responds by repaying loans or depositing additional collateral, thus restoring optimal utilization. Below optimal utilization, the controller will dissipate error, decaying interest rates until the market removes collateral or takes out loans, restoring optimal utilization. The resulting interest rate at optimal utilization can be sustained as the controller maintains the accumulated error, and does not adjust rates further until utilization is altered. Subsequently, the risks associated with high utilization are mitigated, so optimal utilization can be increased, increasing overall revenue without introducing additional risk to the protocol or to lenders. The strategy is sustainable, predictable, and requires no human intervention to correct for changing market conditions and risk appetites. The proposed control system architecture and equations are generally typical of PID control systems and have been tailored to suit the use case. The proposed interest rate strategy is modeled as per Figure \ref{fig:pid-irm}.

In this strategy, the input signal is converted to Utilization Error by subtracting the Optimal Utilization (Equation \ref{eqError}).

\begin{equation}
  U_{error} = U - U_{optimal},
  \label{eqError}
\end{equation}

where $U_{error}$ is the utilization error, $U$ is the utilization, and $U_{optimal}$ is the optimal utilization.

In Figure \ref{fig:pid-irm}, Optimal Utilization was considered to be 0.5, but can initialize anywhere from 0 to 1, (0\% to 100\% Utilization). $U_{error}$ is normalized such that the utilization range $[0, U_{o}]$ is scaled to $[-1, 0]$ and the range $[U_{o}, 1]$ is scaled to $[0, 1]$. The result is a more generalized strategy that can be applied to markets of varying optimal utilization without affecting the configuration of the controller. Utilization Error then passes through the controller.

Equation \ref{eqControllerError} defines the output of the controller as a modified version of the utilization error, which accounts for the proportional, derivative, and integral components calculated by the controller.

\begin{equation}
  U_{\text{controllerError}} = U_{P} + U_{I} + U_{D},
  \label{eqControllerError}
\end{equation}

where $U_{controllerError}$ is the modified utilization error (controller output), and $U_{P}$, $U_{I}$, and $U_{D}$ are the proportional, integral, and derivative components of utilization, respectively.

Due to the cumulative nature of the integrator component, the modified utilization error signal will transition from being sensitive to the proportional component, to being predominantly driven by the integrator component. In essence, the controller modifies the utilization error to account for the size of the error, how long the error has been sustained, and how quickly the error was incurred.

This allows the controller to increase the modified signal when utilization exceeds optimal, or decrease the signal when utilization drops to below optimal, while compensating additionally for short term risks causing fast utilization changes.

The controller error signal passes through a Transfer Function, which follows similar design guidelines to the conventional piecewise functions, but is vastly different. The new Transfer Function is derived from first principles and because the input signal is modified by the controller, it need only provide a baseline interest rate to account for low accumulated error utilization at optimal and maximum utilizations, before error has enough time to accumulate. This serves as a deterrent and not a primary defense mechanism, because the controller will always adapt to market conditions.

\section{Transfer Function}

Since the controller modifies rates in accordance with market activities, the transfer function is characteristic of the desired interest rate relationship, without being overly prescriptive. It is designed by considering guidelines and deriving a simple equation. The transfer function should:

\begin{itemize}
  \item Pass through (0,0), such that the interest rate is programmed to be 0 when utilization is 0.
  \item Pass through some reasonable baseline interest rate at optimal utilization $(U_{o}, r_{o})$. This serves to scale yield if the market size scales early on, before any error can accumulate, and anchors the mechanism to optimal utilization. Over time, it is expected that the interest rate at $U_{o}$~will deviate from $r_{o}$~based on market risk appetite.
  \item Tends towards a large baseline rate as utilization approaches 1, such that complete utilization is mitigated by a large proportional gain.
\end{itemize}

The transfer function is defined in Equation \ref{eqTf}:

\begin{equation}
  r = m \cdot (\frac{U_{\text{controllerError}} + 1}{2})^{n},
  \label{eqTf}
\end{equation}

where $r$ is the interest rate, [$m$] is a scaling factor that allows for a large interest to be implemented when utilization is high and the integrator term has not had time to accumulate, and [$n$] is a scaling factor that allows for $(U_{o}, r_{o})$ to be implemented by solving for $m$ at the required utilization and rate. Since the controller error is normalized to a range of $[-1, 1]$, the transfer function base is further normalized to a range of $[0, 1]$, to ensure the exponent works as intended.

The result is an easily configurable transfer function that can be adapted to stable markets as well as volatile markets with varying liquidity conditions and protocol risk profiles.

\section{Controller Architecture}

The controller features three components; the proportion component, the integral component, and the derivative component. Each component implements a gain factor, which can be tuned, and is discussed later.

\subsection{Proportional Component}

Equation \ref{eqP} defines the proportional component, which simply takes utilization and applies a proportional gain factor that can be tuned.

\begin{equation}
  U_{P} = K_{P} \cdot U_{\text{error}},
  \label{eqP}
\end{equation}

where [$K_{P}$] is the proportional gain coefficient.

\begin{figure}[!ht]
  \centering
  \includegraphics[width=1\linewidth]{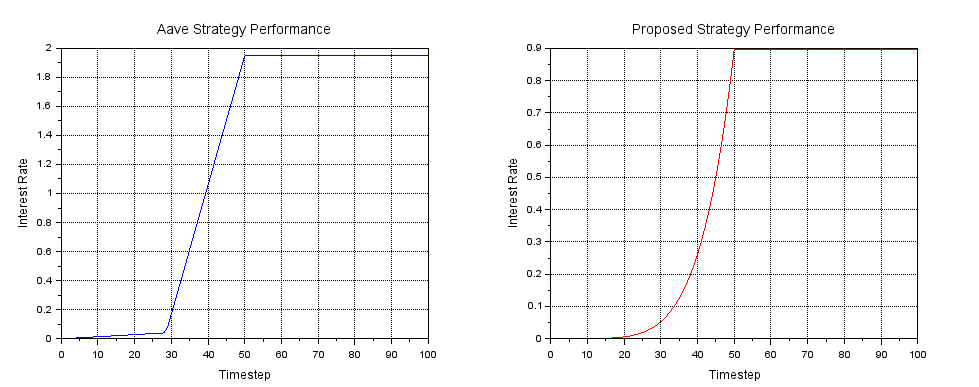}
  \caption{\label{fig:aaveP}Representation of proportional-only strategy performance.}
\end{figure}

Since the Aave IRM is previously modeled, it serves as a good benchmark for comparison, ignoring market response for reasons afforementioned. Figure \ref{fig:aaveP} compares theoretical strategy performance with utilization climbing linearly to 0.8 over 50 timesteps and maintaining $U = 0.8$ for another 50 timesteps, where optimal utilization is 0.5.

Key differences in performance are the evident `kink' in the Aave strategy, when utilization crosses 0.45, which is not present in the continuous curve transfer function, and that the baseline rate at 80\% utilization from the transfer function is around 90\% compared to Aave's approximately 190\%. Clearly, implementing a proportional component alone is not sufficient to manage high utilization risks.

\subsection{Integral Component}

The integral component performs a time series analysis of the market's utilization to generate an accumulated time-weighted error (Equation \ref{eqI}). This is achieved by multiplying the utilization error signal each time it is reported by the time since it was last reported. This rolling calculation is mathematically efficient and can operate effectively ad infinitum. It accumulates when error is positive and dissipates when error is negative, and does so proportionally to the size and sustained time of the error. Accumulated error is then multiplied by an integral gain factor that can be tuned.

\begin{equation}
  U_{I} = K_{I} \cdot \sum_{i=1}^{n} U_{\text{error}, i} \cdot (t_{i} - t_{i-1}),
  \label{eqI}
\end{equation}

where $K_{I}$ is the integral gain coefficient, $U_{error, i}$ is the utilization error at which the $i^{th}$ utilization error is recorded. $t_{i}$ is the time at which the $i^{th}$ utilization is recorded, and $t_{i-1}$ is the previous time at which utilization was recorded.

The integral gain factor can be tuned to grow interest rates appropriately over time. In Figure \ref{fig:aaveI} it has been tuned to demonstrate the effects within the same 100 timesteps:

\begin{figure}[!ht]
  \centering
  \includegraphics[width=1\linewidth]{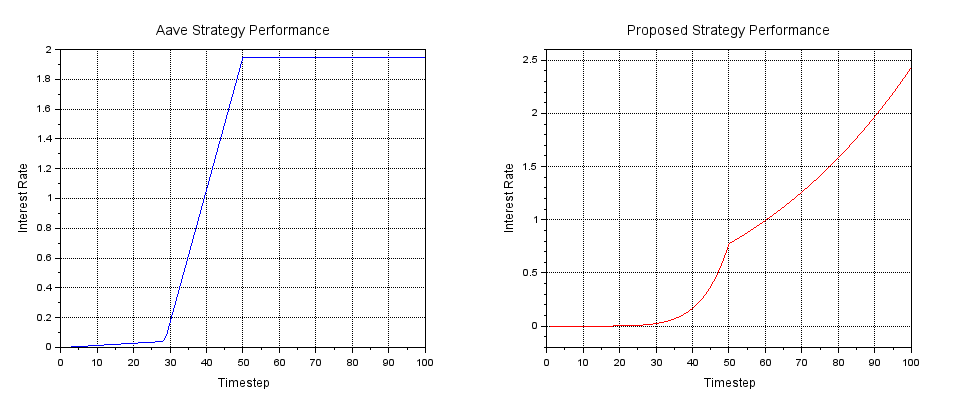}
  \caption{\label{fig:aaveI}Representation of PI strategy performance.}
\end{figure}

In this example, negative error accumulates while utilization is below optimal, and accumulates to the positive when utilization remains above optimal. When utilization reaches 80\% at timestep 50, the interest rate is approximately 80\%, and since 80\% utilization is sustained for the next 50 timesteps, interest rates grow to approximately 240\%. It can be seen that while the Aave strategy has no ability to influence the market beyond its equation rate, the proposed strategy will continue to persuade the market to return to optimal utilization by increasing rates, resulting in either position adjustments or liquidations. If low utilization is sustained, the system can decay rates in the same fashion, making the market more competitive until utilization increases.

If utilization sits below optimal for an extended period of time, there is potential for it to accumulate a large negative value. Should utilization then spike above optimal, the integral value will greatly suppress the proportional component and result in dangerously low interest rates. To mitigate this, a limit condition is programmed such that when the utilization rate is above optimal, the minimum integrator error is $U_{I, min} = -0.5 \cdot U_{P}$. This limits the effect of a negative integrator component on a positive proportional component.

\subsection{Derivative Component}

The derivative component performs another time series analysis of utilization by calculating the gradient over a specified lookback time (Equation \ref{eqTwce}). For this, a time-weighted cumulative error parameter is stored.

\begin{equation}
  \text{TWCE}_{i} = \sum_{i=1}^{n} U_{\text{error}, i} \cdot (t_{i} - t_{i-1}),
  \label{eqTwce}
\end{equation}

where $TWCE_{i}$ is the TWCE at which the $i^{th}$ TWCE is recorded.

After defining a period, the timestamp of each update can be compared to a delayed TWCE. If at least one period has passed, the delayed TWCE and associated timestamp overwrite the previous TWCE and timestamp, and the current TWCE and timestamp overwrite the delayed TWCE and timestamp. The result is two TWCEs that store the cumulative error one period ago, and now, allowing for the derivative calculation presented in Equation \ref{eqD}.

This mitigates the effects of very sudden utilization changes resulting from large borrows/ repayments, which would subsequently make interest rates excessively sensitive to market actions. This is multiplied by a derivative gain factor, which can be tuned, and results in the derivative error.

\begin{equation}
  U_{D} = K_{D} \cdot \frac{TWCE_{delayed} - TWCE_{previous}}{t_{delayed} - t_{previous}},
  \label{eqD}
\end{equation}

where $K_{D}$ is the derivative gain coefficient, $TWCE_{delayed}$ is the most recent TWCE recorded before the current value, and is only updated after at least one period, and $TWCE_{previous}$ is the $TWCE$ recorded at least one period before $TWCE_{delayed}$, and allows for the derivative calculation. $t_{delayed}$ and $t_{previous}$ are the respective timestamps.

The derivative component allows the system to compensate for short term (commensurate with the specified lookback time) market actions, to curb rapid deviations from optimal utilization, protecting the protocol and lenders.

\begin{figure}[!ht]
  \centering
  \includegraphics[width=1\linewidth]{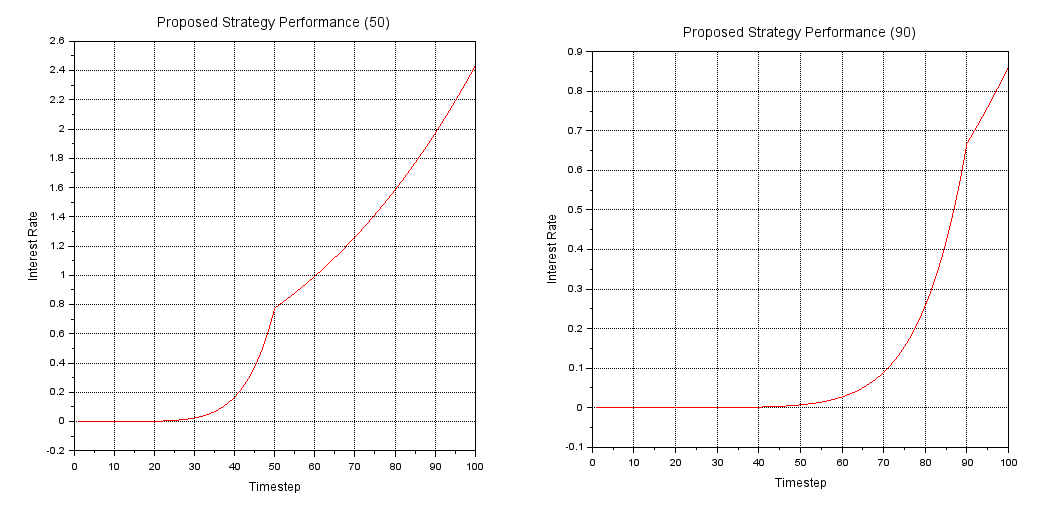}
  \caption{\label{fig:aaveD}Representation of PID strategy performance.}
\end{figure}

In order to demonstrate the effectiveness of the derivative component, it is beneficial to compare utilization increasing to 80\% over 50 timesteps, versus 90 timesteps, before maintaining utilization, as presented in Figure \ref{fig:aaveD}.

On the left, utilization climbs linearly to 80\% over 50 timesteps, and on the right it does the same over 90 timesteps. Since the gradient is lower in the 90 timestep case, the derivative component is reduced. Therefore the short-term compensation is reduced, and the interest rate at the inflection point ($U = 0.8$) is approximately 65\%. In the 50 timestep case, the gradient is higher, increasing the compensation, resulting in $U=0.8$ rates of approximately 80\%. In both cases, when utilization is sustained at 80\%, the derivative component influencing the interest rate decays while the integral component accumulates, growing interest rates further. The result is a strategy that is responsive in both short and long time frames, and will continually adapt to market conditions.

\section{Implementation}

There is no requirement for off-chain infrastructure to support this strategy. Since money markets typically deploy an interest rate strategy smart contract, this architecture simply replaces it.

The arithmetics employed are within the scope of most smart contract programming languages and most of the calculations are computed in memory, reducing gas requirements. Derivative gain architecture may be gas intensive so its use should be considered against the intended network infrastructure. In most cases, it is anticipated that interest rates should not be highly sensitive to derivative gain, and its use is not strictly needed.

Contract architecture has been developed and tested separately and compared in production privately. Some information may be released in future publications.

\section{Conclusion}

The control system described in this paper enables a passively adaptive interest rate strategy that can grow interest rates in high risk environments and decay interest rates in low risk environments, ensuring that the system remains competitive and without taking on additional protocol risk. Exact controller configurations have been omitted due to the sensitive nature of the performance of the strategy. As always, security remains the priority, and any contracts derived from this research should be thoroughly tested and audited before implementation.

\printbibliography

@article{morphoPaper,
  title   = {Agents' behavior and Interest Rate Model Optimization in DEFI lending},
  doi     = {10.2139/ssrn.4802776},
  journal = {SSRN Electronic Journal},
  author  = {Bertucci, Charles and Bertucci, Louis and Gontier Delaunay, Mathis and Gueant, Olivier and Lesbre, Matthieu},
  year    = {2024},
  month   = {4}
}

@misc{aaveIrm,
  title   = {Borrow interest rate: AAVE v3: Risk},
  url     = {https://docs.aave.com/risk/liquidity-risk/borrow-interest-rate},
  journal = {Borrow Interest Rate | Aave V3},
  author  = {Aave},
  year    = {2023},
  month   = {10}
}

@misc{aaveModel,
  title   = {Aave Interest Rate Curve Model},
  url     = {https://medium.com/@zeroalpha/aave-interest-rate-curve-model-9366a69fd2e8},
  journal = {Medium},
  author  = {Yuvi},
  year    = {2023},
  month   = {5}
}

@misc{aaveWhitepaper,
  title   = {Aave Protocol Whitepaper V1.0},
  url     = {https://github.com/aave/aave-protocol/blob/master/docs/Aave_Protocol_Whitepaper_v1_0.pdf},
  journal = {Github},
  author  = {Aave},
  year    = {2020},
  month   = {1}
}

@misc{ajnaWhitepaper,
  title   = {AJNA PROTOCOL: Automated Lending Markets Whitepaper},
  url     = {https://www.ajna.finance/pdf/Ajna_Protocol_Whitepaper_01-11-2024.pdf},
  journal = {Ajna Finance},
  author  = {Ajna},
  year    = {2023},
  month   = {12}
}

@misc{morphoDocs,
  title={Adaptive curve IRM description},
  url={https://docs.morpho.org/contracts/morpho-blue/reference/irm/adaptative-curve-irm/description},
  journal={Morpho Docs},
  author={Morpho},
  year={2024},
  month={4}
}

\end{document}